# Quantitative measurement of birefringence in transparent films across the visible spectrum

Aaron D. Slepkov

Physics & Astronomy, Trent University, Peterborough, Canada

aaronslepkov@trentu.ca

*Abstract*—Common transparent polymer films such as cellophane and household tape are frequently used as examples of birefringent materials in textbooks and classroom demonstrations. Qualitatively, birefringence is often demonstrated by layering such films between crossed-polarizers. In this work, we describe an inexpensive experimental setup for the quantitative measurement of birefringence in common household films, suitable for senior high school or undergraduate labs. Whereas traditional approaches for polarization-based classroom experiments typically use monochromatic laser light, we encourage the combined use of an incoherent incandescent light source and portable spectrometer. In addition, we demonstrate how any concomitant thin-film interference effects can be used to separately measure the optical thickness in the most heterogeneous and uniform films. Such measurement can then be used as an independent experimental confirmation of either the film's index of refraction or its thickness, given knowledge of the other. In an effort to provide examples for the data analysis procedures as well as to investigate a range of materials, we measure the birefringence across the visible spectrum of six common household polymer films, including thin kitchen wrap, cellophane, gift basket film, and common adhesive tapes.

## INTRODUCTION

Optical birefringence is a material property pertaining to a difference in index of refraction that depends on the polarization and propagation direction of light. It is an important topic of study in optics, as birefringence is the central phenomenon used for many important technological applications such as LCD screen displays,[1] modern 3D glasses,[2] and microscopy.[3] For centuries, calcite has remained the quintessential example of a birefringent material.[4] As such, the canonical demonstration of birefringence continues to be the display of double-refraction in calcite; largely because its birefringence is so large, but also because the observation of double-refraction does not require any preparation of the polarization state of the light. However, while birefringence is at the heart of double-refraction,[4] the latter is a challenging concept to describe, and its demonstration can be lacklustre. Beyond calcite, birefringence in polymers and other "plastics" has also become a mainstay textbook example of birefringence. In particular, the fact that household films such as kitchen wrap and adhesive tape can act as retarders (or *waveplates*) remains a common feature in polarization chapters of optics textbooks.[5] Beyond the enticing visual effects of viewing bright tape between crossed polarizers,[6] the mathematical and quantitative details of such birefringence are also important and easily accessible via simple modern lab equipment.[7]

Over the years there have been numerous investigations of birefringence in transparent polymer films.[7–17] Most of these investigations measure the mean index of refraction $n$ or the maximum difference in index between perpendicular polarizations, (i.e., the birefringence $\Delta n$) at single or discrete wavelengths,[8–10, 15–17] where the wavelengths are often chosen by the availability of laser sources.[7] In fact, both detailed quantitative measurements and classroom laboratory demonstrations of polarization-dependent optical phenomena typically involve narrowband (or *monochromatic*) sources.[7, 10, 26, 17–19] It is difficult to discern whether the ubiquity of using lasers for lab demonstrations is pedagogically driven by a desire to keep measurements conceptually restricted to single wavelengths, or simply by a presumption that a coherent light source is needed. Nonetheless, there may be broader pedagogical advantages to conducting conceptually simple but technologically sophisticated lab experiments involving broadband (i.e., "white") and incoherent light; experiments much like those used historically by the likes of Newton, Bernoulli, Brewster, Young, Biot, Rayleigh, and others.

In this letter we describe a straightforward approach to quantify the birefringence of transparent tape and other (quasi-)uniaxial transparent films across the visible spectrum. With a simple setup that includes cheap sheet polarizers, standard lab optics, a portable USB spectrometer, and an incandescent light source, this experiment is ideal for junior-level university/college instructional labs and ambitious senior high school physics classrooms. Throughout this article we will highlight a variety of tricks and traps, knowledge of which may improve the quality of experimental classroom results. Finally, we will present the quantitative measurements of broadband birefringence for a variety of common birefringent polymer films, including adhesive tapes, gift-basket wrap, cellophane, and polyethylene kitchen wrap.

## THEORY

*I. Simple Polarimetry and birefringence*

A quantitative measurement of film birefringence can be obtained by simple polarimetry.[7, 10] In the most general terms, a birefringent sample, sandwiched between a pair of polarizers, alters the amount of light that exits the polarization gate as a function of the orientation of the film within the gate—irrespective of the relative orientation of the polarizers. The transmission of light through the gate depends on the optical wavelength, and thus a collection of the transmitted spectrum offers a straightforward measurement of the birefringent action of the film.

Consider an experimental setup consisting of light passing through an initial polarizer on the way to a birefringent sample, and then onto a second polarizer designated as the *analyzer*. The first polarizer prepares the incident light into a linear polarization state, irrespective of the pre-polarizer polarization state of the light. The birefringent sample alters this linear polarization for any sample orientation where either optical axis isn't exactly aligned along the polarization axis of the first polarizer. If we consider the sample rotated by an angle $\theta$ with respect to the first polarizer's polarization axis, the sample induces a phase retardance between orthogonal polarization components as each travels at a slightly different speed. This *retardance* (in radians) is given by $\delta = \frac{2\pi}{\lambda} d |n_1(\lambda) - n_2(\lambda)|$, where $d$ is the thickness of the sample, $\lambda$ is the wavelength of the light, and $n_1(\lambda)$ and $n_2(\lambda)$ are (generally, wavelength-dependent) indices of refraction in the sample. The birefringence of the sample is then $\Delta n = |n_1(\lambda) - n_2(\lambda)|$, and is not itself thickness or angle dependent, but is rather a material property of the sample.

Once passing through the sample, the polarization state of the light has been altered such that, in general, it is no longer necessarily linear and no longer predominantly aligned along the axis of the first polarizer. In terms of the dominant polarization angle, the sample will have rotated it by an angle of $2\theta$. Depending on the birefringence, thickness, and orientation of the sample, as well as the wavelength under consideration, the outgoing polarization may be elliptical, circular, or linear.[5] The analyzer filters the light along its polarization axis, transmitting an intensity that depends on the incident wavelength, the retardance induced by the sample, and the sample's orientation. Two experimental arrangements are conceptually useful: In the *closed gate* (CG) arrangement, the polarizer and analyzer are *crossed*, such that no light emerges when the sample induces no retardance (or the sample is removed). In the *open gate* (OG) arrangement, the polarizer and analyzer are co-aligned and all colors are maximally transmitted in the no-retardance condition. The spectral intensity transmitted through the closed gate and open gate arrangements are given by

$$I_{\text{CG}}(\lambda) = I_0(\lambda) \sin^2\left(\frac{\pi}{\lambda} d \Delta n(\lambda)\right) \sin^2(2\theta) \quad (1)$$

and

$$I_{\text{OG}}(\lambda) = I_0(\lambda) \left(1 - \sin^2\left(\frac{\pi}{\lambda} d \Delta n(\lambda)\right) \sin^2(2\theta)\right), \quad (2)$$

respectively, where $I_0(\lambda)$ is the incident spectral intensity passing through the first polarizer.

Analysis of these expressions shows that in either arrangement the transmission is a double modulation controlled by the rotation angle of the sample relative to the first polarizer and the retardance provided by the sample. Furthermore, collection of this signal with a spectrometer can yield information about the birefringence at each wavelength, thereby providing a measurement of dispersion in $\Delta n$,[7] rather than assuming a-priori that the birefringence itself is a constant across the visible spectrum. Experimentally, it is most convenient to vary the sample orientation rather than somehow vary the sample thickness or birefringence. Thus, for a given sample, the retardance is constant at each wavelength, and the signal modulation is controlled via rotation of the sample in the plane perpendicular to the propagation axis (i.e., varying $\theta$). A key experimental condition is attained when the film's birefringent axes are set at 45° to the polarizer's axis. In this case, maximum retardance is experienced by the light and the equation for transmission is simplified. This *maximum action* condition yields a normalized transmission of

$$T_{\text{CG}}^{45°}(\lambda) = \left(\frac{I^{45°}(\lambda)}{I_0(\lambda)}\right)_{\text{CG}} = \sin^2\left(\frac{\pi}{\lambda} d \Delta n(\lambda)\right) \quad (3)$$

and

$$T_{\text{OG}}^{45°}(\lambda) = \left(\frac{I^{45°}(\lambda)}{I_0(\lambda)}\right)_{\text{OG}} = \cos^2\left(\frac{\pi}{\lambda} d \Delta n(\lambda)\right). \quad (4)$$

*II. Etalon effects and optical path length*

In some ways, a precise measurement of $\Delta n$ is easier to make than that of the individual indices $n_1$ and $n_2$. The birefringence of polymers ranges widely between values of zero and ~0.015, whereas the index of refraction of most transparent polymer films only differs slightly between 1.45 and 1.6.[20] Thus, estimating the index to be 1.5 is likely adequate for most purposes. Nonetheless, one likely wishes to measure the index of refraction concurrently with a measurement of film birefringence. Fortunately, thin and transparent samples such as those under consideration often display so-called *etalon effects*—a sort of thin-film interface transmission signal that can be used to estimate the index of refraction to a reasonable degree of accuracy.[10, 21]

The amount of light reflected from a dielectric surface in air depends on the index of refraction of the material and the angle of incidence.[5] For fully transparent samples with no absorption, the light that isn't reflected is transmitted into the film. The transmitted light then travels to the back surface, at which point it partially reflects inwards and partially transmits out of the sample. Thus, light in the sample is free to bounce back and forth between the two interfaces, setting up constructive/destructive interference that depends on the wavelength of light and the optical path length of the film, given by *nd*. The resulting interference fringes can be observed both spatially and spectrally. Spectrally, the progression of constructive and destructive interference fringes present as a regular modulation in the transmitted spectrum.[22] Experimentally, the depth of modulation in the spectral fringes varies considerably depending on the surface quality and parallelism of the sample, and the spatial uniformity of the incident beam among other considerations. However, if any fringes are observed, the wavelength separation of consecutive transmission maxima or minima proves to be a useful ruler for measuring the index of refraction (or alternately, the thickness) of the thin film. This separation is called the *free spectral range* of the film (or etalon), and is approximately given by

$$\Delta\lambda_{\text{FSR}} = \frac{\lambda_0^2}{2nd}, \quad (5)$$

where $\lambda_0$ is the wavelength at the midpoint between the fringes.[22] Since both $\lambda_0$ and $\Delta\lambda_{\text{FSR}}$ can be measured with relatively good precision, an analysis of etalon effects present in the transmission signal obtained from the polarimeter can provide an independent measurement of *nd*.

## MATERIALS AND METHOD

The equipment needed for establishing a simple polarimetry gate is simple and relatively inexpensive, with a total cost of approximately US$5,500 for the components shown in Fig. 1. For a detailed parts-list see Appendix B. The most expensive components are the USB spectrometer and its associated computer, along with the rotation stages for the sample holder and analyzer. The use of a simple incandescent halogen floodlight means that the experiment described here is less costly than most other common teaching lab experiments that require a laser source.

In its most basic arrangement, the polarimetry gate comprises an incandescent light source, a polarizer mounted on a fixed axis, a rotatable sample holder, a second polarizer on a rotatable mount, and a spectrometer outfitted with a multimode optical collection fiber. The polarizers should be suitable for covering a significant proportion of the visible spectrum. We have found that most inexpensive sheet polarizers are adequate for the task. The cheapest film polarizers provide reasonable polarization purity to span the wavelength range 500 nm – 700 nm. Thicker polarizer sheets (sold as "laminated") cover the wavelength range 430 nm – 760 nm, and at under US$30 per sheet are highly recommended for these experiments. Professional-grade calcite polarizers can offer improved polarization purity across a broader spectral range. For the experiments presented here, we used a combination of a cube polarizer as the first polarizer and laminated sheet polarizer as the analyzer. The analyzer was simply affixed to a rotatable lens mount via double-sided tape. In general, film polarizers may be used for both polarizers. Beyond the basic elements of the

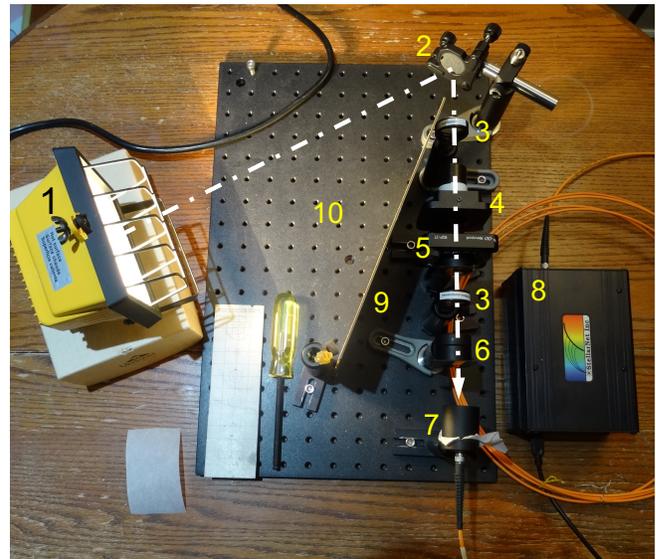

**FIGURE 1.** *Photograph of layout of polarimetric gate for measuring birefringence in polymer films. (1) Incandescent halogen spotlight as light source. (2) 1" silver routing mirror. (3) 20-cm focal length biconcave lenses. (4) ½"-aperture cube polarizer. (5) 1" aperture rotation mount affixed with birefringent sample. (6) sheet polarizer acting as analyzer affixed to a 1" rotatable lens mount. (7) multimode fiber receiving mount with a sheet of diffusing tissue paper stretched between lens tubes. An extended lens tube provides additional rejection of stray light (8) USB spectrometer with multimode fiber input controlled by a tablet computer (not shown). (9) Cardboard sheet acting as a light barrier against direct illumination from light source. (10) Optical mounting breadboard. The dashed white arrow indicates the optical path through polarimeter to detector.*

gate, adjustable mirrors, simple lenses, and optomechanical components on an optical breadboard are useful for better optical alignment and repeatable positioning.

Figure 1 displays a marked-up photograph of our experimental setup. The light source is a simple halogen incandescent spotlight. A halogen-bulb source is preferred for its brightness, which also permits it to be positioned at a distance from the other optical elements. As seen in the figure, we can place the light source as close as ~30 cm from the other optical elements, but we've also had success with the bulb placed at distances greater than 1 m. Instead of pointing the source directly down the axis of the polarization gate, we position it pointing away from the detector, and then direct the incident light down the axis of the gate using a single 1-inch mirror. Stray light that doesn't pass through all the elements of the gate is adequately avoided by use of such routing, along with simple light barriers (i.e., cardboard) and lens tubes around the fiber collector. We have found that the illumination intensity passing through the gate is **not** a limiting factor, whereas subtle alignment variations can lead to significant variability in the spectrum obtained through the direct illumination of the optical fibre. This is largely because of the mismatch between the diameter of the collection fiber and that of the illumination spot that passes through the gate. The best means of making uniform and robust collection is to diffuse the light through thin paper so that the detector is sampling a consistent uniform area. We have found that this is easily implemented by placing a taut swatch of tissue paper (1- or 2-ply) within 2-5 mm in front of the collection fiber, as can be seen in Fig. 1. While this diffuser reduces the overall signal by nearly two orders of magnitude, it provides excellent repeatability and uniformity to the collected intensity.

Birefringent film samples are mounted onto a 1" lens tube screwed into a rotation mount using either double-sided tape or, in the case of tape samples, their own adhesive. The rotation mount has 2º demarcations, which allow for detailed control of sample orientation. For experimental confirmations of Eqs. (1) and (2), transmission spectra are collected as a function of sample angle in 2º increments across any span of 90º. There should be four *no-action* and four *max-action* sample orientations found across each 360º sample rotation in either open/closed gate configuration. The *no-action* conditions occur when the incident polarization is aligned with either of the two birefringence axes of the sample. This condition is easiest to establish by closing the polarization gate and rotating the sample until minimum transmission is observed across the widest wavelength span. Once this condition is found, the analyzer may be rotated by 90º to establish the *no-action* condition in the open gate. The *max-action* condition is then most readily established by rotating the sample 45º from the no-action condition. Figure 2 shows raw spectra for the no-action closed-gate, no-action open gate, and 45º (max action) open gate conditions for a single-layer sample of Scotch® 371 packaging adhesive tape.

Birefringent polymer film samples can be easily found among commonplace household items. Most (but not all!) transparent adhesive tape products are birefringent. These include most packaging and gift-wrap tapes, both of the 19-mm and 48-mm width varieties. While we present results for several 3M Scotch brand tape, the equivalent products from other companies such as Grand & Toy, Duck, Staples, and Paperway are equally birefringent. Both true cellophane and biaxially-oriented polypropylene (BOPP) film are highly birefringent. Traditional cellophane gift-basket film is increasingly difficult to find, and most 'cello' film being sold now is BOPP. A roll purchased from a local dollar store provides a particularly inexpensive birefringent sample. As we show in this work, the birefringence of cellophane and BOPP are not equivalent. We have procured thin square sheets of true cellophane marketed as candy wrappers from Foilman (www.foilman.com). Cellophane and BOPP film can be cut to size and mounted onto a lens tube using thin strips of double-sided tape. Finally, all kitchen 'plastic wrap' that we have tested is birefringent, including both Saran and Glad brands. These samples are very thin and somewhat stretchable. To mount kitchen-wrap for study, we form a taut layer across a lens tube and secure it in place with an elastic band, forming a kind of drum. It's important to take care not to stretch the film as this alters the sample thickness and impacts both the measurement of birefringence and etalon fringes.

The sample rotation angle $\theta$ is defined with respect to the optical axes of the samples. In practice, the orientation of optical axes must be first established experimentally, as described above. For samples that come on a roll, a natural reference direction is established by the direction along the length of the film. Interestingly, we find that various birefringent samples are manufactured such that the roll axis and optical axis do not coincide. For example, this optical-axis-offset can be as high as 30° in (19-mm) rolls of stationary tape, while for other samples such as cellophane sheets and basket wrap there is effectively no offset. All packaging tape varieties showed a roll-axis offset of 9°±3°.

Knowledge of film thickness is necessary for a quantitative estimate of sample birefringence. Non-adhesive sample thickness was measured mechanically: Cellophane and BOPP film were stacked several layers thick and measured with calipers. Kitchen cling wrap thickness was estimated from reported roll length and measured roll stack thickness. An accurate measurement of the birefringence 'backing film' thickness in adhesive tape samples was precluded by the variable and unknown layer thickness of the glue. For those samples, we use the nominal thickness reported by the manufacturer.[23]

## RESULTS AND DISCUSSION

### I. Transmission Analysis

The key to obtaining a normalized transmission spectrum lies in the proper collection of the transmission spectra through both a closed gate and open gate. In each case, the sample is in place but aligned to a no-action orientation. As an example, Fig. 2 presents the raw and normalized transmission curves for a single layer of Scotch packaging tape. In practice, transmission curves were collected for each sample at 2º increments across 100º rotation span. This practice was done as a check of the protocol, and is ultimately unnecessary because the maximum-action condition for all wavelengths occurs when the film is rotated such that $\theta = 45°$. As can be seen in Fig. 2, the closed-gate transmission curve (with sample oriented to $\theta = 0°$) becomes a control measurement for the wavelength-dependent polarization extinction capabilities of the polarizer-analyzer pair. From this curve, we see that our polarimeter gate provides strong extinction between 350 nm and 750 nm, with rapidly dwindling extinction abilities above 750 nm and complete loss of polarization at 960 nm. The no-action open-gate transmission represents a reference measurement for the incident spectrum—that is, it's a proxy for $I_0(\lambda)$. To obtain a normalized curve for $T_{OG}^{45°}(\lambda)$, the raw open-gate transmission with the tape in the $\theta = 45°$ position must be divided by the open-gate transmission, with the no-action closed-gate transmission signal first subtracted from both. The resulting

April 2022

normalized transmission curve is shown in the bottom panel of Fig. 2.

The most noteworthy feature of the open-gate transmission of the Scotch packing tape sample shown in Fig. 2 is the extinction of transmission at wavelengths between 725 nm – 750 nm. This represents a 90º polarization rotation by the sample, *thus signifying its action as a half-waveplate for these wavelengths*. Two oriented layers of this tape would thus comprise a half-waveplate at 360 nm – 375 nm. Unfortunately, both of these wavelength ranges are just outside of the visible spectrum, and thus this tape is not particularly useful as a demonstration of a half-waveplate in the visible regime. Furthermore (and contrary to some reports in the literature[7, 15, 16]), while the six samples studied in this work display a range of birefringence and thicknesses, no sample was found to act as a half-waveplate between 450 nm and 700 nm.

*II. Retardance inversion errors*

The retrieval of birefringence data from open/closed-gate transmission requires the inversion of sine and cosine functions. A simple inversion of Eqs. (3) and (4) will only yield retardance values within a restricted domain of 0 to $\pi/2$. For many low-birefringence samples such as kitchen wrap, the retardance is sufficiently low across the visible spectrum that a direct retrieval of $\Delta n$ works well. However, many samples provide retardance greater than $\pi/2$, and thus a direct inversion of these equations may result in error. There are four possible values of retardance between $\delta = 0$ and $2\pi$ that satisfy any observed value of transmission. The inability to distinguish between these four values is likely one reason why even single-wavelength (i.e., laser-based) polarimetry measurements of birefringence are rarely found in the experimental physics curriculum. However, with a broadband wavelength-dependent measurement, it is possible to discern which of the four domains are being spanned, based on the shape of the retrieved birefringence—at least for most birefringent thin-film samples. Figure 3 shows an example of transmission curves for a simulated sample with sufficiently large birefringence (or thickness) to span several retardance domains across the visible spectrum. Whereas the simulated birefringence is relatively uniform as a function of wavelength, a mindless retrieval of $\Delta n(\lambda)$ based on the transmission curves leads to erroneous results and misleading interpretation. As seen in Fig. 3, only the open-gate transmission in a wavelength range that corresponds to $\delta < \pi/2$ gives the correct value of the birefringence from direct inversion.

At shorter wavelengths, $\Delta n$ is seen to either rise or fall quasi-linearly, forming sharp kinks at wavelengths that correspond to $\left(\frac{1}{2}\right)^i 4d\Delta n$, where $i$ is an integer. The shape of these curves can be used to discern a 'corrected' value of the birefringence. For example, a positive (quasi-)linear slope with wavelength obtained in an open gate arrangement implies that the retardance is between $\pi/2$ and $\pi$. For a closed gate arrangement, a rising positive slope implies that the retardance is between 0 and $\pi/2$, or between $\pi$ and $2\pi$, as shown in Fig. 3. Conversely, a negative-going slope in an open gate arrangement implies a retardance between $\pi$ and $2\pi$, and in a closed gate arrangement it implies a retardance between 0 and $\pi/2$. Alternatively, fitting the raw transmission spectra shown in Fig. 3 (or, for example, that in the bottom panel of Fig. 2) to Eqs. (3) or (4) can provide correct values for $\Delta n(\lambda)$. However, such fitting algorithms will invariably require substantial sophistication if one allows for the birefringence to vary by wavelength.

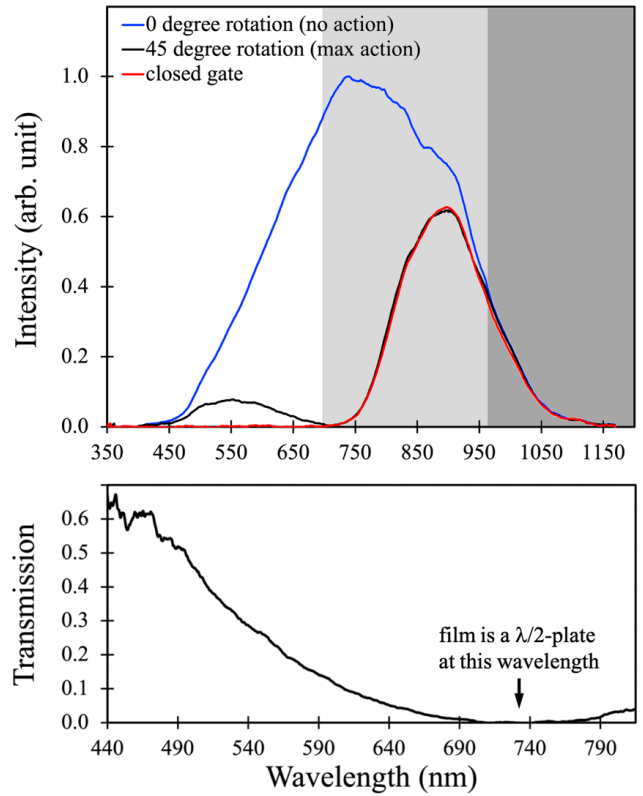

**FIGURE 2.** *Obtaining a transmission spectrum for a single layer of Scotch 371® packing tape oriented at $\theta = 45º$ within an open-gate polarimeter arrangement. (Top panel) Spectra of three gate conditions with sample in place. The no-action condition (blue) is obtained with the sample rotated such that the incident polarization is along one of the optical axes of the film and thus no birefringence is observed. The closed-gate condition (red) is the reference baseline condition that also reflects the extinction spectrum of the polarizer-analyzer combination, showing diminishing polarization purity in the polarization pair above 750 nm (light grey region), and complete loss of polarization above 960 nm (dark grey region). (black) Maximum retardance is effected when the sample is rotated to $\theta = 45º$. (Bottom panel) Transmitted spectrum for the maximum-retardance open-gate condition, represented by Eq. (4). The curve is obtained from the data in the top panel via (black-red)/(blue-red).*

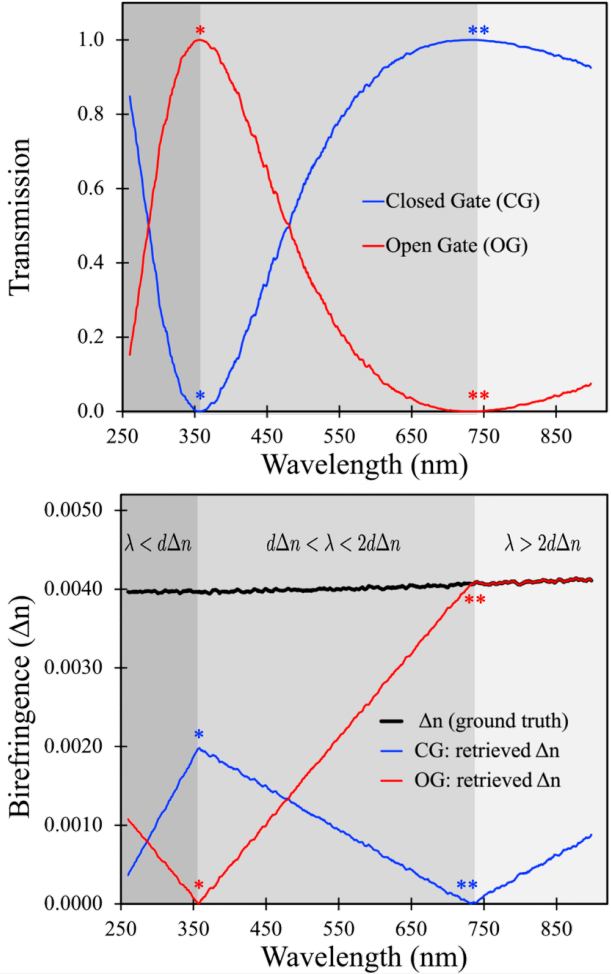

**FIGURE 3.** *Simulated transmission spectra and inversion errors for a birefringence sample with 90-mm thickness oriented at θ = 45° within closed and open gate arrangements. (Top Panel) The transmitted spectra for open- and closed-gate arrangements are mirror images that sum to 1. The open-gate (closed-gate) condition yields maximum (minimum) transmission at wavelengths that correspond to full-waveplate retardance, and yield minimum (maximum) transmission at wavelengths that correspond to half-waveplate retardance. (Bottom Panel) Applying Eqs. (3) and (4) to the spectra in the top panel generates an erroneous curve for the sample birefringence. Only the portion of the transmission spectrum for which wavelengths are above a value of 2dΔn yield the correct birefringence. Proper retrieval of the wavelength-dependent birefringence (i.e., the ground truth) is obtained either with proper fitting of transmission spectra or with manual de-trending of the retrieved birefringence, as described in the main text.*

*III. Etalon resonances*

Several samples displayed etalon fringes in their raw transmission spectrum. For the most part, adhesive tape samples did not display strong fringes, likely because of the inhomogeneity of the adhesive-polymer interface. Non-adhesive samples such as polyethylene kitchen wrap, polypropylene wrap film, and cellophane sheets display prominent etalon fringes that grow and diminish in finesse across the measured wavelength ranges. Figure 4 shows an example of the transmission fringes observed for each of these three samples. According to Eq. (5), a measurement of the wavelength difference between peaks (i.e., the free spectral range, FSR) at any particular wavelength can constitute a measurement of the optical thickness $nd$. As expected, with the three polymers having very similar indexes, the thickness of the film strongly affects the FSR. The kitchen wrap film is considerably thinner than the other two samples, and thus displays fringes that are further apart than the others. In principle, this method could be used to independently measure $n_1$ and $n_2$ of a birefringent film. However, with birefringence below 0.02 (such as found for all samples studied here), differences in the FSR on the order of 0.1 nm would have to be measured. This requirement is well outside of the sampling resolution of most spectrometers—especially of the kind that can be found in teaching labs. Fortunately, this criterion also means that the sample orientation (i.e., $\theta$) at which the spectral fringes are analyzed is unimportant, and the film can be oriented to maximize fringe visibility (as long as normal incidence is maintained).

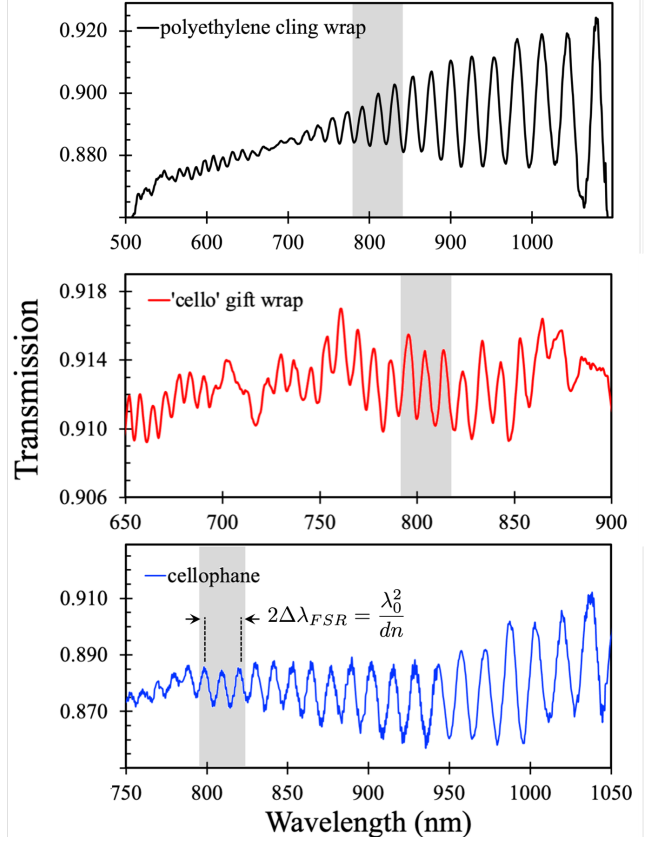

**FIGURE 4.** *Etalon effects in homogeneous thin films at normal incidence. From top to bottom are the normalized transmission spectra for Glad Cling Wrap, Voila 'cello' gift basket film, and Foilman cellophane candy wrapper sheets. The shaded area in each panel spans the two resonances closest to $\lambda_0=800$ nm, which are then used together with Eq. (5) to estimate either the thickness or index of refraction in each of the samples. The noticeably wider free spectral range in the kitchen wrap film results from its appreciably smaller thickness.*

As shown in Fig. 4, all three samples display sufficiently clear fringe patterns in the vicinity of 800 nm to estimate a value of $nd$ for the sample. For each film, we use the value of $nd$ obtained from Fig. 4, together with the mechanically measured film thickness, to obtain an estimated measure of the film's nominal index $n_o$. Such analysis yields values of $1.53 \pm 0.08$, $1.49 \pm 0.06$, and $1.45 \pm 0.06$ for the index of refraction of cling wrap, BOPP 'cello' gift wrap, and cellophane, respectively. The uncertainty in these measurements reflects uncertainties in the thickness measurement, and spectrometer sampling resolution.

*IV. Sample birefringence*

The measured birefringence for six film samples is shown in Fig. 5. In principle, the spectral intensity and polarization purity in our measurements are sufficient to acquire data between 400 nm and 950 nm. In practice, a variety of factors restrict the range over which reliable measurements can be made to 430 nm – 800 nm. The choice between implementing an open-gate or closed-gate measurement is largely arbitrary, as both methods prove reliable. A general guiding approach is as follows: Use the gate configuration that leads to maximum-action transmission that is most distinct from the no-action open-gate transmission. Thus, a preference may be made for closed-gate measurements for samples with either very small retardances, or with retardance approaching integer multiples of $\pi$. Conversely, open-gate measurements are best for samples with retardances closer to multiples of $\pi/2$. The birefringence curves shown in Fig. 5 largely reflect this choice: The smallest retardance samples—Scotch transparent tape and Glad cling wrap—are measured in the closed-gate arrangement, as is the sample of Scotch Storage and Packing tape which yields the largest retardance among all measured samples. On the other hand, Cellophane, polyethylene gift basket wrap, and Scotch (371) Packing Tape, were measured in the open-gate arrangement.

Measurements of birefringence in thin films are often collected at a single wavelength, with the presumption that this material property is constant in spectral regions far from any absorptive resonances. For the most part, the curves shown in Fig. 5 confirm this presumption; that is, the birefringence of the films studied in this work are relatively constant across a wide swath of the visible spectrum. A slight downturn in birefringence is observed at wavelengths below 500 nm, but only in the three samples measured in an open-gate arrangement. The three measurements were collected weeks apart. Thus, this artefact is likely due to some systematic error in the collection of the incident spectrum in the open-gate arrangement. Nonetheless, all samples show essentially constant birefringence between 500 nm – 800 nm.

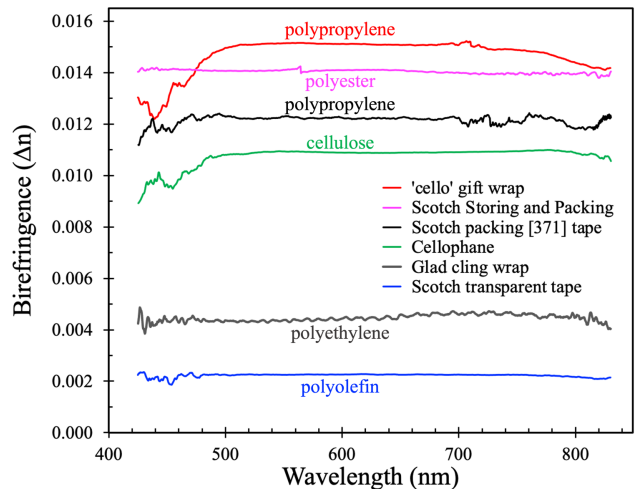

**FIGURE 5.** *Measured birefringence from transparent household films. Three samples were measured with the open-gate arrangement ('cello' gift wrap, Scotch packing [371] tape, and cellophane). The other three samples were measured in the closed gate arrangement. The birefringence for Scotch packing tape (black) is obtained from the transmission data presented in Fig. 1.*

Our selection of samples represents a variety of chemical compositions and manufacturing processes. A summary of sample attributes, including the measured birefringence, reported for the wavelength range of 500 nm–700 nm, is presented in Table 1. The largest birefringence is found in transparent gift basket wrap film that is often sold as 'cello warp', in reference to the traditional cellophane-based films that were ubiquitous decades ago. Nowadays, this film is made of BOPP rather than cellulose. Interestingly, with a measured value of $\Delta n = 0.015 \pm 0.001$, the birefringence of this BOPP film is about 33% larger than the $0.011 \pm 0.001$ measured for the cellophane candy-wrapper film. Scotch 371 packaging tape, also made of BOPP, yields a birefringence of $0.012 \pm 0.002$. This value—while still larger than that for cellophane—is smaller than that for the non-adhesive BOPP basket-wrap film, suggesting that the manufacturing process can significantly affect the resulting birefringence.[26] The slightly thicker packaging tape sold as Scotch Storage and Packaging is made of bidirectionally oriented polyester (rather than polypropylene), and shows a relatively large birefringence of $0.014 \pm 0.001$. At nominally 10-mm thickness, Glad Cling Wrap is the thinnest sample investigated. This polyethylene film displays a relatively small birefringence of $0.0043 \pm 0.0002$. In a twist of irony, we find that standard transparent stationary tape yields the lowest birefringence of the measured samples, with $\Delta n = 0.0023 \pm 0.0009$. Nowadays this tape is made of polyolefin, but historically transparent stationary tape was made of much more birefringent cellophane. In fact, in many parts of the world this product is still known generically as Sellotape. We suspect that when transparent tape was first used for textbook demonstrations of birefringence, the tape was invariably made of cellophane.[5] Compared with other common birefringent materials, the birefringence measured for polymer films is large but unremarkable. By comparison, calcite displays tenfold larger birefringence than the most birefringent BOPP sample, whereas birefringence in quartz is about 20% smaller than that of cellophane.[27] Ice, with reported birefringence of 0.0014 is about half as birefringent as polyolefin tape.[28] Nonetheless, the birefringence of BOPP and cellulose film are sufficiently large to enable their use in polarimetric technologies such as optical displays and "polage" art.[6, 11, 29]

TABLE 1. CHEMICAL, PHYSICAL, AND OPTICAL ATTRIBUTES FOR SIX TRANSPARENT BIREFRINGENT POLYMER FILMS

| Sample | Backing Material | $n_o$ | Thickness, $d$ (μm) | $\Delta n$ (500 nm – 700 nm) | optical axis offset (±3°) |
|---|---|---|---|---|---|
| Foilmam™ Cellophane | Cellophane | 1.47[20] ** <br> 1.45 ± 0.06 * | 20.8 ± 0.1 † | 0.011 ± 0.001 | 0° |
| Voila™ 'cello' wrap | BOPP | 1.51[24] ** <br> 1.49 ± 0.06 * | 23.5 ± 0.2 † | 0.015 ± 0.001 | 0° |
| Scotch™ [371] box-sealing tape | BOPP + adhesive layer | 1.51[24] ** | 30[23] ** | 0.012 ± 0.002 | 9° |
| Glad cling wrap® | Polyethylene | 1.54[21] ** <br> 1.53 ± 0.08 * | 10.9 ± 0.9 † | 0.0043 ± 0.0002 | 18° |
| Scotch™ Storage & Packaging tape | Biaxially-oriented Polyester + adhesive layer | 1.52[25] ** | 40[23] ** | 0.014 ± 0.001 | 9° |
| Scotch™ Transparent tape (gloss finish) | Polyolefin + adhesive layer | 1.5[25] ** | 38[23] ** | 0.0023 ± 0.0009 | 30° |

\*   estimated optically
\*\*  from literature
†   measured mechanically

## SUMMARY

In this work we have described an inexpensive experimental setup for the quantitative measurement of birefringence in thin polymer films, suitable for senior secondary or upper-level college/university instructional labs. Whereas traditional approaches for polarization-based classroom experiments typically use monochromatic laser light, we encourage the combined use of spectroscopic measurement and incoherent unfiltered light sources. There are significant advantages to this approach, including garnering an appreciation for the wavelength dependence of optical retardance, and the ability to estimate the absolute birefringence in samples that experience retardances greater than π/2. In addition, we demonstrate how thin-film interference effects can be used to measure the optical thickness, $nd$, in heterogeneous films of uniform thickness. Such measurement can then be used as an independent experimental confirmation of either the index of refraction or film thickness, given knowledge of the other. In an effort to provide examples for the data analysis procedures as well as to impart novel information not readily available in the literature, we report the measured birefringence in six common household polymer films, including thin kitchen wrap and common adhesive tapes.

## ACKNOWLEDGEMENTS

This work grew out of an International Baccalaureate high school physics project conducted during a global pandemic by Jan (Jack) Francis Beda. I thank Jack for preliminary data collection, for testing early iterations of the experimental setup, and for productive advice on data analysis. Professor Rayf Shiell is gratefully acknowledged for formative manuscript review and guidance.

## APPENDIX A: OBTAINING EXPRESSIONS FOR INTENSITIES TRANSMITTED THROUGH BIREFRINGENT GATES

Derivation of Eqs. (1) and (2) comprise a straightforward and useful exercise that can double as an introduction to Jones Calculus.[30, 31]

We consider transmission through a *closed* birefringence gate to comprise incident incandescent light passing through a polarizer with vertical polarization axis, then passing through a birefringent sample with fast axis aligned $\theta$ degrees from the horizontal, and finally passing through an analyzer with horizontal polarization axis. For an *open* birefringence gate, the sequence is identical except that the analyzer passes vertical polarization. Each of these elements can be represented as a Jones matrix as follow:

Vertically-polarized light field vector after input polarizer:
$$\bar{E}_o(\lambda) = E_o(\lambda) \begin{bmatrix} 0 \\ 1 \end{bmatrix}.$$

Polarizers with vertical (V) and horizontal (H) polarization axes:
$$P_V = \begin{bmatrix} 0 & 0 \\ 0 & 1 \end{bmatrix}; \quad P_H = \begin{bmatrix} 1 & 0 \\ 0 & 0 \end{bmatrix}.$$

The birefringent sample is a linear retarder with fast axis at angle $\theta$ to the horizontal, and a max retardation of $\delta$:[30, 31]

$$R(\theta) = e^{-i\frac{\delta}{2}} \begin{bmatrix} \cos^2\theta + e^{i\delta}\sin^2\theta & (1-e^{i\delta})\cos\theta\sin\theta \\ (1-e^{i\delta})\cos\theta\sin\theta & \sin^2\theta + e^{i\delta}\cos^2\theta \end{bmatrix} = \begin{bmatrix} \cos\frac{\delta}{2} + i\sin\frac{\delta}{2}\cos 2\theta & i\sin\frac{\delta}{2}\sin 2\theta \\ i\sin\frac{\delta}{2}\sin 2\theta & \cos\frac{\delta}{2} - i\sin\frac{\delta}{2}\cos 2\theta \end{bmatrix}.$$

Recalling that Jones matrices are multiplied from output to input, the transmitted light field emerging from a closed gate is given by

$$\bar{E}_{CG}(\lambda) = P_H R(\theta) \bar{E}_o(\lambda)$$
$$= \begin{bmatrix} 1 & 0 \\ 0 & 0 \end{bmatrix} \begin{bmatrix} \cos\frac{\delta}{2} + i\sin\frac{\delta}{2}\cos 2\theta & i\sin\frac{\delta}{2}\sin 2\theta \\ i\sin\frac{\delta}{2}\sin 2\theta & \cos\frac{\delta}{2} - i\sin\frac{\delta}{2}\cos 2\theta \end{bmatrix} \begin{bmatrix} 0 \\ 1 \end{bmatrix} E_o(\lambda)$$
$$= \begin{bmatrix} 1 & 0 \\ 0 & 0 \end{bmatrix} \begin{bmatrix} i\sin\frac{\delta}{2}\sin 2\theta \\ \cos\frac{\delta}{2} - i\sin\frac{\delta}{2}\cos 2\theta \end{bmatrix} E_o(\lambda)$$
$$= E_o(\lambda) \begin{bmatrix} i\sin\frac{\delta}{2}\sin 2\theta \\ 0 \end{bmatrix}.$$

The transmitted spectral intensity is given by $I(\lambda) = \frac{1}{2}\varepsilon_0 c \bar{E}(\lambda) \bar{E}^*(\lambda)$, where the asterisk (*) denotes the complex transpose. Thus, the spectrum of the light transmitted through the closed birefringence gate is

$$I_{CG}(\lambda) = \frac{1}{2}\varepsilon_0 c E_0^2(\lambda) \begin{bmatrix} i\sin\frac{\delta}{2}\sin 2\theta \\ 0 \end{bmatrix} \begin{bmatrix} -i\sin\frac{\delta}{2}\sin 2\theta & 0 \end{bmatrix}$$
$$= \frac{1}{2}\varepsilon_0 c E_0^2(\lambda) \sin^2\frac{\delta}{2} \sin^2 2\theta.$$

Defining the incident spectrum as $I_0(\lambda) = \frac{1}{2}\varepsilon_0 c \bar{E}_0^2(\lambda)$, and recalling that the retardance is given by $\delta = \frac{2\pi}{\lambda}d\Delta n$, one recovers Eq. (1);

$$I_{CG}(\lambda) = I_0(\lambda) \sin^2\left(\frac{\pi}{\lambda}d\Delta n(\lambda)\right) \sin^2(2\theta).$$

Likewise, the transmitted light field emerging from an open gate is given by

$$\bar{E}_{OG}(\lambda) = P_V R(\theta) \bar{E}_o(\lambda)$$
$$= \begin{bmatrix} 0 & 0 \\ 0 & 1 \end{bmatrix} \begin{bmatrix} \cos\frac{\delta}{2} + i\sin\frac{\delta}{2}\cos 2\theta & i\sin\frac{\delta}{2}\sin 2\theta \\ i\sin\frac{\delta}{2}\sin 2\theta & \cos\frac{\delta}{2} - i\sin\frac{\delta}{2}\cos 2\theta \end{bmatrix} \begin{bmatrix} 0 \\ 1 \end{bmatrix} E_o(\lambda)$$
$$= \begin{bmatrix} 0 & 0 \\ 0 & 1 \end{bmatrix} \begin{bmatrix} i\sin\frac{\delta}{2}\sin 2\theta \\ \cos\frac{\delta}{2} - i\sin\frac{\delta}{2}\cos 2\theta \end{bmatrix} E_o(\lambda)$$
$$= E_o(\lambda) \begin{bmatrix} 0 \\ \cos\frac{\delta}{2} - i\sin\frac{\delta}{2}\cos 2\theta \end{bmatrix}.$$

The transmitted spectrum is

$$I_{OG}(\lambda) = I_0(\lambda) \begin{bmatrix} 0 \\ \cos\frac{\delta}{2} - i\sin\frac{\delta}{2}\cos 2\theta \end{bmatrix} \begin{bmatrix} 0 & \cos\frac{\delta}{2} + i\sin\frac{\delta}{2}\cos 2\theta \end{bmatrix}$$
$$= I_0(\lambda) \left( \cos^2\frac{\delta}{2} + \sin^2\frac{\delta}{2} \cos^2 2\theta \right).$$

Using $\cos^2 2\theta = 1 - \sin^2 2\theta$ and rearranging, we recover Eq. (2);

$$I_{OG}(\lambda) = I_0(\lambda) \left( 1 - \sin^2\left(\frac{\pi}{\lambda}d\Delta n(\lambda)\right) \sin^2(2\theta) \right).$$

Finally, note that the two conditions are complementary, such that $I_{CG}(\lambda) + I_{OG}(\lambda) = I_0(\lambda)$.

## APPENDIX B: LIST OF PARTS AND COMMENT ON COSTS

Table 2. provides a list of parts and prices that most closely represents the full experimental setup used for the present work. All optics and optomechanical mounts are sourced from a single vendor (Thorlabs) to best assure compatibility among the items. However, any of a variety optical instrument vendors may be used without loss of functionality. Likewise, the USB spectrometer used for the present work is similar to ones sold by a variety of companies; any of which are likely to be suitable for the kinds of measurements reported here. As presented, the total extended price is approximately US$5,500. Except for sheet polarizers, nothing was purchased specifically for the present study; the equipment was already part of a well-stocked laser lab. Thus, establishing a viable spectroscopically-based polarization gate for the expressed use in a teaching laboratory can be done at considerably less cost, and without significant loss of functionality. The greatest cost savings can be had by using sheet polarizer instead of a calcite polarizer cube, and by purchasing an educational-quality spectrometer instead one of research grade. In toto, the experiments described in this work could be implemented, without significant loss in quality, with a setup costing under US$3,500.

TABLE 2. Parts list and notes for assembling a birefringence-measurement setup

| Item # in Fig. 1 | General description | Parts Used (or equivalent) | Approx. Cost $USD/ea | Notes |
|---|---|---|---|---|
| 1 | Halogen spotlight | NOMA 500W | $40 | Must be incandescent. Gets hot; use on-demand. |
| 2 | Mounted alignment mirror | *PF10-03-P01<br>*KM 100 | $55<br>$40 | 1" silver mirror<br>Kinematic mirror mount |
| 3 | Mounted focusing lenses (×2) | *LB1945-ML | $35 (×2) | 20-cm focal-length biconvex lenses. Other lenses also work well. |
| 4 | 1st polarizer on rotation mount | *GL5<br>*LRM1<br>*SM1PM10 | $700<br>$100<br>$50 | 5-mm Glan Laser Calcite Cube polarizer<br>Rotation mount<br>Polarizer cube adapter |
| 5 | Rotation mound for birefringent sample | *LRM1 | $100 | Sample can be affixed to mount with double-sided adhesive tape. |
| 6 | Analyzer affixed to simple lens mount | ***PF030<br>*LMR1<br>*CM1L03 | $35<br>$15<br>$20 | Laminated sheet polarizer. Price is for 12" × 17". Cut to 2" × 2".<br>Lens mount<br>0.3"-long lens tube |
| 7 | Multimode fibre and collection mount | *M35L01<br>*SM1SMA<br>*CM1L10 | $100<br>$35<br>$20 | 1-mm SMA Large-core "patch cord"<br>Threaded SMA adapter plate<br>1"-long lens tube |
| 8 | USB spectrometer | **BLK-CXR | $3200 | Black Comet spectrometer |
| 9 | Simple light barrier | | | Found cardboard object |
| 10 | Optical breadboard | *MB1218 | $205 | 12" × 18", ½" thick aluminum breadboard |
| N/A | Post holders (×7)<br>Posts (×7)<br>Clamping forks (×7)<br>Post bases (×7) | *PH2<br>*TR3<br>*CF125<br>*BE1 | $9 (×7)<br>$6 (×7)<br>$10 (×7)<br>$10 (×7) | 2" depth post holder<br>3"-long posts<br>Mounting forks<br>1" bases |
| N/A | Spectrometer display and control computer | Repurposed laptop with spectrometer software. | $300 | Often included with spectrometer purchase. |

\*   Thorlabs part number.
\*\*  Stellarnet part number. Recommended substitute: Green Wave educational spectrometer ($2,000)
\*\*\* Manufactured by Aflash Photonics. Distributed via Polarization.com.

# REFERENCES

[1] H. Kawamoto, "The history of liquid-crystal display and its industry," 2012 Third IEEE HISTory of ELectro-technology CONference (HISTELCON), 1-6 (2012).

[2] I. P. Howard, and B. K. Rogers, *Stereoscopic Vision*, (Oxford University Press, 2012).

[3] H.F. Talbot Esq. M.P. F.R.S. XLIV. "Experiments on light," The London, Edinburgh, and Dublin Philosophical Magazine and Journal of Science **5** (29), 321-334 (1834).

[4] E. Bartholin, "Experimenta crystalli islandici disdiaclastici quibus mira & insolita refractio detegitur," Copenhagen 1669. English translation: "Experiments with the double refracting Iceland crystal which led to the discovery of a marvelous and strange refraction," tr. by Werner Brandt. Westtown, Pa., 1959.

[5] E. Hecht, *Optics*, 4th ed. (Addison-Wesley, 2002).

[6] A. D. Slepkov, "Painting in polarization," Submitted to Am. J. Phys. (2022) [forthcoming]

[7] A. Beléndez, E. Fernández, J. Francés, and C. Neipp, "Birefringence of cellotape: Jones representation and experimental analysis," Eur. J. Phys. **31** 551 (2010).

[8] K. Iizuka, "Cellophane as a half-wave plate and its use for converting a laptop computer screen into a three-dimensional display," Rev. Sci. Instrum. **74**, 3636 (2003)

[9] P. Velasquez, M. del Mar Sánchez-López, I. Moreno, D. Puerto, and F. Mateos, "Interference birefringent filters fabricated with low cost commercial polymers," Am. J. Phys. **73**, 4 (2005).

[10] M. A. Blanco, M. Yuste, and C. Carreras, "Undergraduate experiment designed to show the proportionality between the phase difference and the thickness of a uniaxial crystal," Am. J. Phys. **65**, 784–787 (1997).

[11] A. R. Wood, "Polage: A language of kinetic color," MSc. Thesis, Syracuse University, August 1981.

[12] Z. Bashir, S. Bandyopadhyay, R. Kummetha, and J. Lohmeijer, "On the development of uniaxially-oriented PET tapes for weaving, woven tape fabrics, and their applications," Polym Eng Sci, **59**: E120-E132. (2019).

[13] V. B. Gupta, and S. Kumar, "Intrinsic birefringence of poly(ethylene terephthalate)," J. Polym. Sci. Polym. Phys. Ed., **17**, 1307-1315 (1979).

[14] S. J. Edwards, and A. J. Langley, "On Producing Colours Using Birefringence Property of Transparent, Colourless Stretched Cellophane," Leonardo, **14** (3) 187-190 (1981).

[15] D. Kinyua, G. Rurimo, P. Karimi, S. Maina, and C. Ominde, "Interferometry Analysis of Cellophane Birefringence," Optics and Photonics Journal, **3**, 337-341 (2013).

[16] M. Ortiz-Gutiérrez, A. Olivares-Pérez, and V. Sánchez-Villicaña, "Cellophane film as half wave retarder of wide spectrum," Opt. Mater. **17**, 395-400 (2001).

[17] A. Khanra, and B Raychaudhuri, "Cellotape Birefringent Filter: Some New Demonstrations," Optics and Photonics Journal, **6**, 139-145 (2016).

[18] B. Wang, and T. C. Oakberg, "A new instrument for measuring both the magnitude and angle of low level linear birefringence," Rev. Sci. Instrum. **70**, 3847-3854 (1999).

[19] G. E. Jellison, and F. A. Modine, "Two-modulator generalized ellipsometry: experiment and calibration," Appl. Opt. **36**, 8184-8189 (1997).

[20] S. N. Kasarova, N. G. Sultanova, C. D. Ivanov, and I. D. Nikolov, "Analysis of the dispersion of optical plastic materials," Opt. Mater., **29**, 1481–1490, (2007).

[21] B. Šantić, "Measurement of the refractive index and thickness of a transparent film from the shift of the interference pattern due to the sample rotation," *Thin Solid Films*, **518** 3619-3624 (2010).

[22] Wikipedia contributors. Fabry–Pérot interferometer. Wikipedia, The Free Encyclopedia. October 19, 2021. Available at: https://en.wikipedia.org/w/index.php?title=Fabry%E2%80%93P%C3%A9rot_interferometer&oldid=1050753557. Accessed January 3, 2022.

[23] "3M Packaging Solution Guide," 3M Closure and Masking Systems Division, 70-0710-0432-2 Rev April 2019.

[24] F. Hashimoto, "Biaxially oriented polypropylene film and method for its production," U.S. Patent 4,405,775, issued September 20, 1983.

[25] B. S. Mitchell, *An Introduction to Materials Engineering and Science: For Chemical and Materials Engineers*, Appendix 9, (John Wiley & Sons, Inc, 2004).

[26] P. Dias, A. Hiltner, E. Baer, J. Van Dun, H. Chen, and S.P. Chum, "Structure and Properties of Biaxially Oriented Polypropylenes (BOPP)," *ANTEC 2006*, 2660-2664 (2006)

[27] R. N. Smartt, and W. H. Steel, "Birefringence of Quartz and Calcite," J. Opt. Soc. Am. **49**, 710-712 (1959)

[28] P. V. Hobbs, *Ice physics*, (Oxford University Press, 2010).

[29] A. Wood-Comarow, "Labyrinthe De Lumière," Leonardo **24**, 84-86 (1991).

[30] Wikipedia contributors. Jones Calculus. Wikipedia, The Free Encyclopedia. 13 January 2022. Available at: https://en.wikipedia.org/wiki/Jones_calculus. Accessed April 20, 2022.

[31] E. Collett, *Field Guide to Polarization*, (SPIE Press, 2005).